\documentclass{IEEEtran}

\usepackage[margin=1in]{geometry}
\usepackage{graphicx}
\usepackage[style=ieee,backend=biber]{biblatex}
\usepackage{hyperref}
\hypersetup{
  colorlinks=true,
  linkcolor=black,
  filecolor=magenta,
  urlcolor=blue,
  citecolor=black,
  }
\usepackage{breakurl}
\begin{document}

\title{Apple Silicon Performance in Scientific Computing}
\author{Connor Kenyon and Collin Capano,\\\textit{Center for Scientific Computation and Data Science Research}
\thanks{This article was submitted for peer-review on \today. This work was supported in part by NSF Grant No. DMS-19127165.}
\thanks{Connor Kenyon is a member of the Physics Department and the Center for Scientific Computing and Data Science Research at the University of Massachusetts Dartmouth, North Dartmouth, MA 02747 USA (e-mail: ckenyon@umassd.edu).}
\thanks{Dr. Collin Capano is member of the Physics Department and the Center for Scientific Computing and Data Science Research at the University of Massachusetts Dartmouth, North Dartmouth, MA 02747 USA, and a senior scientist with the Max-Planck-Institut f{\"u}r Gravitationsphysik (Albert-Einstein-Institut), D-30167 Hannover, Germany (e-mail: ccapano@umassd.edu).}}
\date{}
\maketitle

\begin{abstract}
    With the release of the Apple Silicon System-on-a-Chip processors, and the impressive performance shown in general use by both the M1 and M1 Ultra, the potential use for Apple Silicon processors in scientific computing is explored. Both the M1 and M1 Ultra are compared to current state-of-the-art data-center GPUs, including an NVIDIA V100 with PCIe, an NVIDIA V100 with NVLink, and an NVIDIA A100 with PCIe. The scientific performance is measured using the Scalable Heterogeneous Computing (SHOC) benchmark suite using OpenCL benchmarks. We find that both M1 processors out perform the GPUs in all benchmarks.
\end{abstract}

\begin{IEEEkeywords}
    Scientific Computing, Embedded Devices, Accelerators, Prallel Computing, Supercomputing, GPU, GPGPU, OpenCL, SHOC, Physics
\end{IEEEkeywords}

\section{Introduction}

\IEEEPARstart{R}{ecent} trends in high performance computing (HPC) have shown a significant shift in types of processors and architectures being utilized in clusters. There is a transition from discrete CPU and GPU systems to systems that include heterogeneous architectures. This shift is most visible within the Top500 supercomputer list, in which the top 3 as of June 2022 --- Frontier, Fugaku, and Lumi --- all take advantage of high core-count heterogeneous processors \cite{top500}. The main benefit of heterogeneous architectures is the ability to utilize shared memory in order to minimize or entirely eliminate data transfer times. This is done through zero-copy algorithms, or algorithms that allow for memory access from both the CPU and GPU without requiring data transfer \cite{song2012performance}. Removing communication bottlenecks in this way has been shown to improve performance by as much as a factor of 30 in data-center GPUs \cite{kenyon2019overcoming}.

With the introduction of the Apple line of System-on-a-Chip (SoC) processors, Apple Silicon, there is now an additional manufacturer of heterogeneous architectures \cite{apple2020}. The competitive processors from Apple include their M1 series, as well as the newly announced M2 series. The first M1 processor was released in November 2020, followed by
the M1 Pro and M1 Max in October of 2021, and the M1 Ultra in March of 2022 \cite{haslam2022}. Initial CPU and GPU benchmarks showed significant promise in the overall performance capabilities of the M1 \cite{android2022, M1UltraCPU, M1UltraGPU}. With their relatively low cost, low power consumption, and the general trend of HPC toward SoC processors, the potential for using these new Apple Silicon processors in scientific computing applications is manifold. 

This paper is organized as follows: in Section~\ref{sec:hardware} specific details about the tested hardware are discussed; in Section~\ref{sec:benchmarks} the benchmarks are introduced and explained; results are shown in Section~\ref{sec:results}; concluding remarks are in Section~\ref{sec:conclusions}. 

\section{Hardware Details}
\label{sec:hardware}

The hardware being compared in this work represents current state-of-the-art HPC GPUs alongside two of the Apple silicon processors. This allows for a comparison to be made between performance that would be obtainable on a standard compute node in an HPC cluster and a Mac computer featuring one of the Apple Silicon processors.

For data-center GPUs, we use two NVIDIA Tesla V100s \cite{V100_specs} --- one with PCIe interconnect and one with the SXM2 interconnect --- along with an NVIDIA Tesla A100 \cite{A100_specs}. These represent the previous and current generations of NVIDIA data-center GPUs, respectively. They have been a staple in HPC clusters since their releases. The specifications for these are outlined in Table \ref{tab:gpu_specs}.

We benchmark two Apple Silicon processors: the first-generation M1 \cite{m1_specs} and the M1 Ultra \cite{m1ultra_specs}, the latter of which is the the most recent and top-performing processor in the M1 series. Specs for both are shown in Table \ref{tab:m1_specs}.

\begin{table}[htbp!]
    \centering
    \caption{Hardware details for NVIDIA V100 and A100 GPUs.}
    \label{tab:gpu_specs}
    \begin{tabular}{c|c|c}
        &   V100 & A100 \\\hline
        Memory &  32 GB HBM2 & 80 GB HBM2 \\\hline
        Memory Bandwidth & 900 GB/s & 1,935 GB/s \\\hline
        TDP & 300W & 300W \\\hline
        Interconnect & PCIe 3.0 / NVLink & PCIe 4.0 \\
    \end{tabular}
\end{table}

\begin{table}[htbp!]
    \centering
    \caption{Hardware details for Apple M1 and M1 Ultra.}
    \label{tab:m1_specs}
    \begin{tabular}{c|c|c}
        &   M1 & M1 Ultra \\\hline
        Memory &  16 GB LPDDR4 & 32 GB LPDDR5 \\\hline
        Memory Bandwidth & 4,226 MB/s & 800 GB/s \\\hline
        TDP & 15 W & 60 W \\\hline
        Interconnect & N/A & N/A \\
    \end{tabular}
\end{table}

\section{SHOC Benchmarks}
\label{sec:benchmarks}

The SHOC benchmark suite is an open source collection of benchmark programs
selected to measure the performance of a variety of computing devices with a
focus on multi-core performance tests. The benchmarks are available in
multi-core MPI, OpenCL, and CUDA, which provides support for most modern
heterogeneous computing hardware \cite{danalis2010scalable}. In this work, 
we benchmark the Nvidia GPUs using the CUDA benchmarks, and the Apple M1s 
using the OpenCL benchmarks.

To subdivide the particular benchmarks, there are three different performance
measurement levels: level 0 measures the low level behavior; level 1 measures
higher level behavior with commonly used algorithms; level 2 measures
realistic application kernels. Specifically, level 0 benchmarks include:
\begin{itemize}
    \item Bus Speed Download;
    \item Bus Speed Readback;
    \item Device Memory;
    \item Kernel Compile;
    \item Max Flops;
    \item Queue Delay.
\end{itemize}
Level 1 benchmarks are:
\begin{itemize}
      \item Breadth-First Search (BFS);
      \item Fast Fourier Transform (FFT);
      \item Molecular Dynamics (MD);
      \item MD5Hash;
      \item Reduction;
      \item Single-Precision Matrix Multiplication (SGEMM);
      \item Scan;
      \item Sort;
      \item Sparse matrix-vector multiplication (SPMV);
      \item Stencil2D;
      \item Triad.
\end{itemize}
Level 2 benchmarks include Quality Threshold clustering (QTC) and S3D.

\section{Results}
\label{sec:results}

The performance of all processors being compared in various benchmarks allows for a broad comparison of the capabilities of each device for scientific computing. We show the results of SHOC benchmarks in single precision for all five processors. Single precision benchmarks are selected because of the lack of double precision capability in the Apple M1s' GPU.

For the level 0 Max FLOPs benchmark, we find peak single precision performance is lowest with the V100 with PCIe ($1.40\times10^4$~GFLOPs), followed by the V100 with NVLink ($1.55\times10^{4}$~GFLOPs). The A100 outperforms both V100s at $1.93\times10^{4}$~GFLOPs. However, both the Apple M1 and M1 Ultra dwarf this figure, at $2.97\times 10^8$ and $6.56 \times 10^8$ GLOPs peak performance, respectively. The reported performance for the M1s, while impressive, seems significantly larger than plausible.

\begin{figure}[htbp!]
  \centering
  \includegraphics[width=\columnwidth]{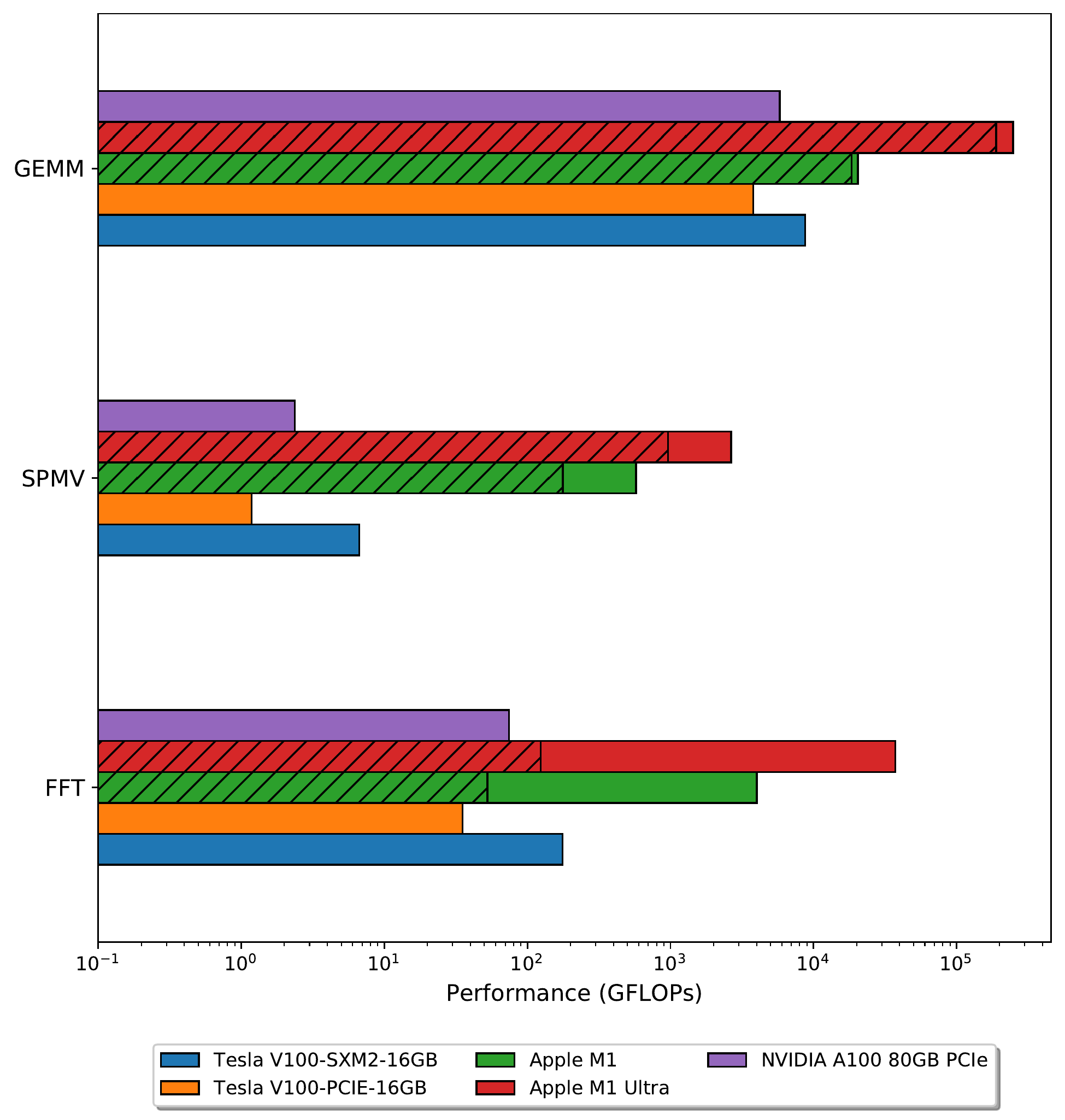}
  \caption{Set of SHOC Level 1 benchmark results showing Generic Matrix Multiplication (GEMM), Sparse Matrix Vector Multiplication (SPMV), and Fast Fourier Transform (FFT) on all devices. Hatched lines indicate performance with shared memory that includes transfer time.}
  \label{fig:l1}
\end{figure}

\begin{figure}[htbp!]
  \centering
  \includegraphics[width=\columnwidth]{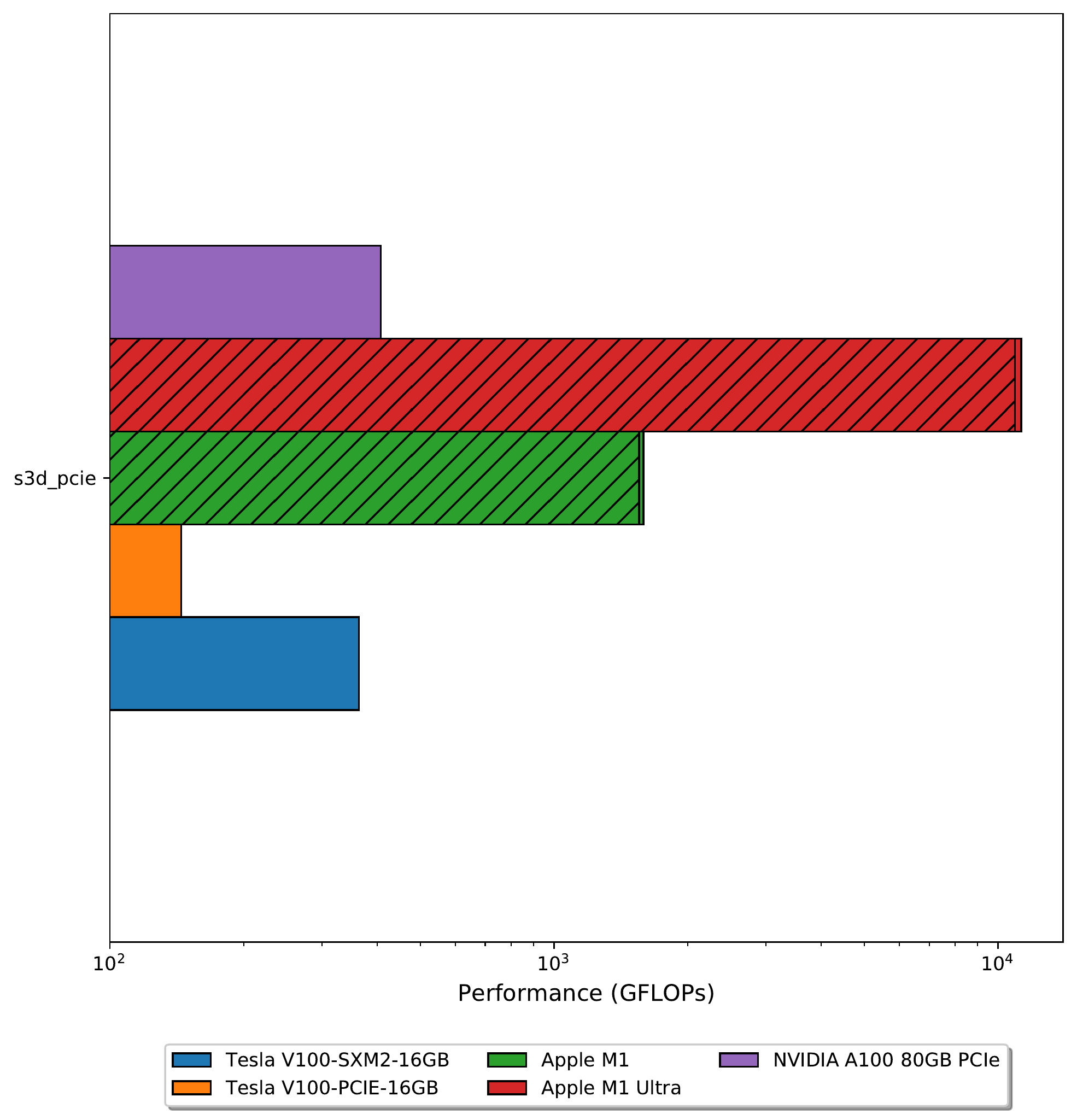}
  \caption{Set of SHOC Level 2 benchmark results showing only S3D on all devices.}
  \label{fig:l2}
\end{figure}

\begin{figure}[htbp!]
    \centering
    \includegraphics[width=\columnwidth]{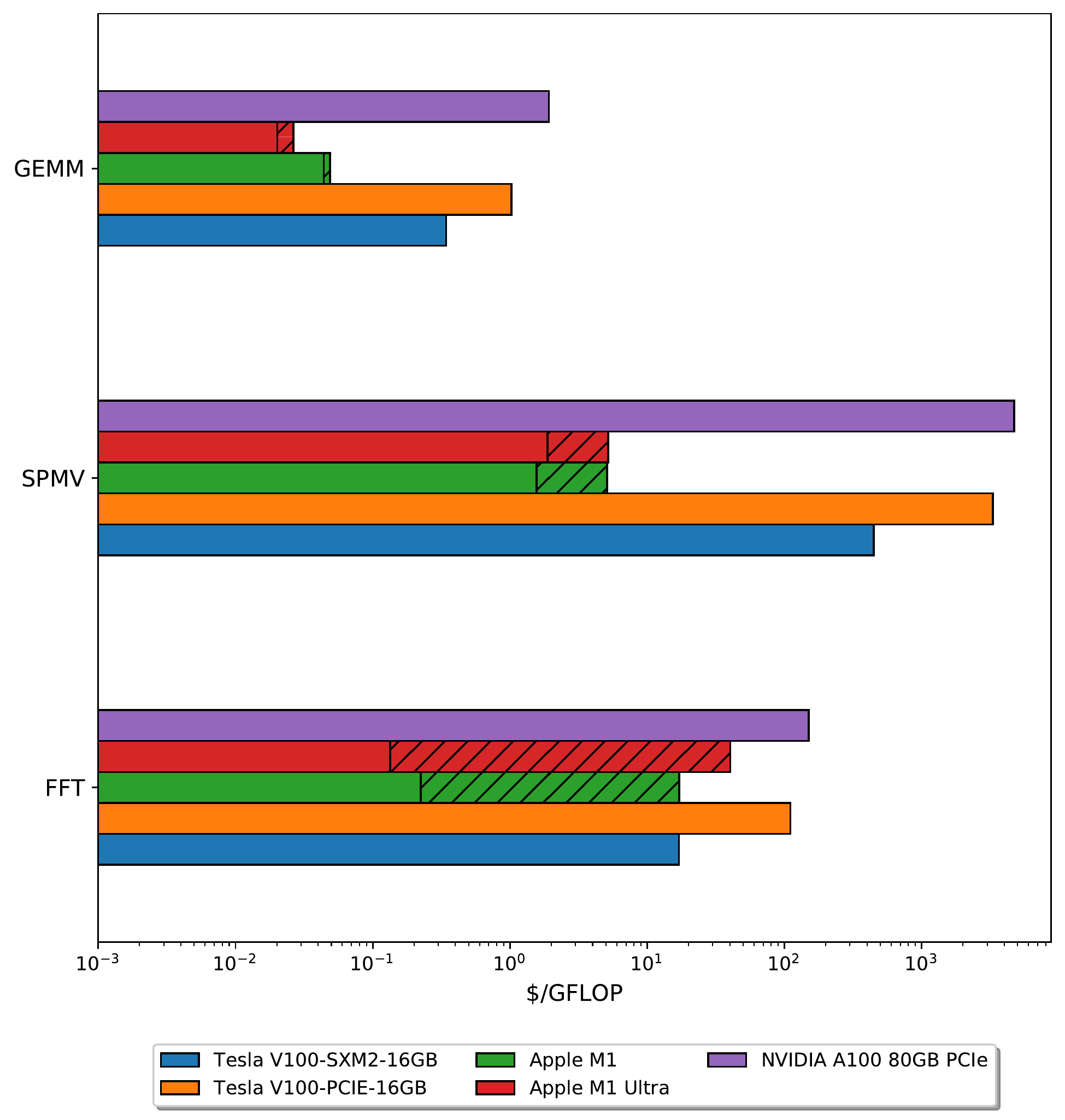}
    \caption{The cost per GFLOP of performance in each processor (lower cost per GFLOP is better), shown for matrix multiplication (GEMM), sparse matrix-vector multiplication (SPMV), and fast Fourier transform (FFT). Hatched lines indicate that the performance when data transfer time is included for a processor with shared memory available.}
    \label{fig:cost}
\end{figure}

In Figure~\ref{fig:l1}, a subset of the level 1 SHOC benchmarks are shown. This subset includes Matrix Multiplication (GEMM), Sparse matrix-vector multiplication (SPMV), and Fast Fourier Transform (FFT).
The performance shown in Figure \ref{fig:l1} includes hatched bars for the M1 and M1 Ultra because they are able to utilize shared memory. The hatched portion indicates the performance with transfer time, while the portion without hatches shows the performance with zero-copy algorithms being used. The real performance when optimally using the hardware is reflected by the bars without hatches. These hatched bars are used similarly in Figures \ref{fig:l2} and \ref{fig:cost}

The NVIDIA GPUs performance varies widely between each of the SHOC level 1 benchmarks shown in Figure \ref{fig:l1}, with the V100 with NVLink consistently outperforming both the V100 and the A100 with PCIe. The performance is strongest on matrix multiplication, where the NVIDIA GPUs all perform competitively with the Apple M1, while still falling an order of magnitude behind the M1 Ultra. For FFTs, both the Apple M1 and M1 Ultra maintain approximately an order of magnitude of separation in performance, while outperforming all three GPUs. In SPMV this difference is seen even more dramatically, with  a separation of two orders of magnitude between the V100 with NVLink and the M1, with the M1 Ultra performing beyond that.

Benchmark results for S3D are shown in Figure \ref{fig:l2}. This is the largest benchmark problem featured in the SHOC benchmarks, and reflects the realistic performance of the processors when used with typical scientific code. Figure \ref{fig:l2} compares all processors, including the NVIDIA V100 and A100 as well as the Apple M1 and M1 Ultra.

The results for S3D shown in Figure \ref{fig:l2} show a similar pattern to the SPMV benchmark seen in Figure \ref{fig:l1}. The most notable change is that the A100 performs slightly better than the V100 with NVLink, but performance still falls significantly behind both the M1 and the M1 Ultra. The performance of both Apple processors maintains a significant margin above all three Nvidia GPUs. The difference between the hatched and unhatched bars for the M1 and M1 Ultra show nearly no difference. This reflects a heavily operations-bound curve for S3D, meaning shared memory independently provides no significant benefit to this algorithm.

A pertinent question is how much each processor costs relative to its performance. Comparing costs between the GPUs and the M1s is complicated by the fact that the M1 processors can only be purchased as a part of a complete Mac desktop (which comes with storage) while the Nvidia GPUs can be purchased on their own (without storage). However, as a rough approximation we use the cheapest available Mac Mini with the M1 tested here, and the cheapest available Mac Studio with the M1 Ultra tested here. For the Nvidia GPUs, current available GPU prices are used. Specifically, we estimate the price of each processor as: 
\begin{itemize}
    \item Apple M1: \$899.99 \cite{macmini}
    \item Apple M1 Ultra: \$4,999.99 \cite{macstudio}
    \item Nvidia V100 PCIe: \$3,899.00 \cite{NeweggV100}
    \item Nvidia V100 NVLink: \$2,990.00 \cite{NeweggV100SXM}
    \item Nvidia A100: \$11,157.19 \cite{NeweggA100}
\end{itemize}
The cost per GFLOP is shown in Figure \ref{fig:cost} (here, smaller is better).

The cost per GFLOP shown in Figure \ref{fig:cost} has the M1 and M1 Ultra performing comparably in each benchmark. In GEMM, the M1 and M1 Ultra cost an order of magnitude less per GFLOP. For SPMV and FFT the difference is more substantial, with the difference between the M1 and M1 Ultra and the Nvidia GPUs increasing to over two orders of magnitude. The Nvidia GPUs maintain a relatively fixed difference in price per GFLOP relative to one another, with the V100 with NVLink consistently the lowest cost per GFLOP, followed by the V100 with PCIe, and then the A100 at the largest cost per GFLOP in all benchmarks shown. A significant highlight of these benchmarks is that by optimally utilizing shared memory with data transfer-dominated algorithms, such as FFT, there is a two order of magnitude decrease in price per GFLOP of performance. 

\section{Conclusions}
\label{sec:conclusions}

Recent trends in high performance computing have shown a significant shift towards heterogeneous system architectures, often including ARM processors as an integral component of the system. Previously, this market has been dominated by Nvidia and AMD. The Apple silicon processors introduce new competition.

While there are some features that the Apple M1 and M1 Ultra lack, such as double precision GPU computing capabilities, the single precision performance shows an incredible amount of promise. Though currently only available in Mac computers and not sold individually, these processors show impressive performance that is applicable to many machine learning or single precision-dominated research in scientific computing. Even when considering cost of purchase, the price per GFLOP of an M1 or M1 Ultra comes at a fraction of the cost of a data-center GPU and accompanying server node.

Heterogeneous system architectures are an extremely promising path forward in high performance computing. Despite the lack of double precision GPU computing in the Apple M1 series, the performance is promising and may see use in single-precision applications in research computing. While there is no sign of these becoming available to use in clusters, they will be able to provide robust single precision computing on a workstation.

The authors would like to thank Gaurav Khanna, Scott Field, and Geoffrey Cowles for useful comments and discussions. This work was supported in part by NSF Grant No. DMS-19127165.

\nocite{*}
\printbibliography

@inproceedings{danalis2010scalable,
  title={The scalable heterogeneous computing (SHOC) benchmark suite},
  author={Danalis, Anthony and Marin, Gabriel and McCurdy, Collin and Meredith, Jeremy S and Roth, Philip C and Spafford, Kyle and Tipparaju, Vinod and Vetter, Jeffrey S},
  booktitle={Proceedings of the 3rd Workshop on General-Purpose Computation on Graphics Processing Units},
  pages={63--74},
  year={2010}
}

@misc{top500, 
  title={Top500 June 2022},
  url={https://top500.org/lists/top500/2022/06/}, 
  journal={TOP500}
}

@misc{apple2020, 
   title={Apple announces Mac transition to Apple Silicon},
   url={https://www.apple.com/newsroom/2020/06/apple-announces-mac-transition-to-apple-silicon/},
   year={2020},
   month={Jun}
 }

@misc{haslam2022, 
  title={M1 chip guide: M1, M1 Pro, M1 Max and M1 Ultra},
   url={https://www.macworld.com/article/676321/m1-pro-max-ultra-guide.html}, 
   author={Haslam, Karen}, 
   year={2022}, 
   month={Jun}
 }

@misc{android2022, 
  title={Apple M1 tested: Performance benchmarks and thermal throttling, explained},
  url={https://www.androidauthority.com/apple-m1-test-benchmark-performance-thermal-1185988/},
  year={2022}, 
  month={Apr}
}

@inproceedings{kenyon2019overcoming,
  title={Overcoming Limitations of GPGPU-Computing in Scientific Applications},
  author={Kenyon, Connor and Volkema, Glenn and Khanna, Gaurav},
  booktitle={2019 IEEE High Performance Extreme Computing Conference (HPEC)},
  pages={1--9},
  year={2019},
  organization={IEEE}
}

@misc{V100_specs, 
  title={Nvidia formally announces pcie TESLA V100: Available later this year},
  url={https://www.anandtech.com/show/11559/nvidia-formally-announces-pcie-tesla-v100-available-later-this-year},
  journal={RSS}, 
  publisher={AnandTech}, 
  author={Oh, Nate}, 
  year={2017}, 
  month={Jun}
}

@misc{A100_specs, 
  title={Nvidia A100 gpus power the modern data center},
  url={https://www.nvidia.com/en-us/data-center/a100/}, 
  journal={NVIDIA}
}

@misc{m1_specs, 
  title={Apple M1 chip: Specs, performance, everything we know},
  url={https://www.tomshardware.com/news/Apple-M1-Chip-Everything-We-Know}, 
  journal={Tom's Hardware}, 
  publisher={Tom's Hardware}, 
  author={Ehrhardt, Michelle}, 
  year={2021}, 
  month={Jun}
}

@misc{m1ultra_specs, 
  title={Apple unveils M1 Ultra, the world's most powerful chip for a personal computer},
  url={https://www.apple.com/newsroom/2022/03/apple-unveils-m1-ultra-the-worlds-most-powerful-chip-for-a-personal-computer/}, 
  journal={Apple Newsroom}, 
  year={2022}, 
  month={Jun}
}

@article{song2012performance,
  title={Performance review of zero copy techniques},
  author={Song, Jia and Alves-Foss, Jim},
  journal={International Journal of Computer Science and Security (IJCSS)},
  volume={6},
  number={4},
  pages={256},
  year={2012},
  publisher={Citeseer}
}

@misc{macmini, 
  title={Mac Mini},
  url={https://www.apple.com/shop/buy-mac/mac-mini/apple-m1-chip-with-8-core-cpu-and-8-core-gpu-256gb},
}

@misc{macstudio, 
  title={Mac Studio},
  url={https://www.apple.com/shop/buy-mac/mac-studio/20-core-cpu-48-core-gpu-32-core-neural-engine-64gb-memory-1tb},
}

@misc{M1UltraCPU,
  title={M1 Ultra CPU GeekBench},
  url={https://browser.geekbench.com/v5/cpu/14664498},
}

@misc{M1UltraGPU,
  title={M1 Ultra GPU GeekBench},
  url={https://browser.geekbench.com/v5/compute/4757238},
}

@misc{NeweggV100,
  title={Newegg Nvidia V100 PCIe},
  url={https://www.newegg.com/nvidia-900-2g500-0000-000/p/1DH-00RJ-00001}
}

@misc{NeweggV100SXM,
  title={Newegg Nvidia V100 NVLink},
  url={https://www.newegg.com/p/1FT-0004-002M1}
}

@misc{NeweggA100,
  title={Newegg Nvidia A100 PCIe},
  url={https://www.newegg.com/nvidia-900-21001-0000-000/p/N82E16814132090}
}

\onecolumn

\appendices

\section{NVIDIA V100 PCIe SHOC Single Precision Output}

\begin{verbatim}
    result for bspeed_download:                 12.3233 GB/sec
    result for bspeed_readback:                 13.1718 GB/sec
    result for maxspflops:                    14039.2000 GFLOPS
    result for gmem_readbw:                    934.3160 GB/s
    result for gmem_readbw_strided:            404.3400 GB/s
    result for gmem_writebw:                   725.9060 GB/s
    result for gmem_writebw_strided:            18.6136 GB/s
    result for lmem_readbw:                   10286.9000 GB/s
    result for lmem_writebw:                  11825.5000 GB/s
    result for tex_readbw:                    1528.0900 GB/sec
    result for ocl_kernel:                       0.0019 sec
    result for ocl_queue:                        0.0034 ms
    result for bfs:                             15.4394 GB/s
    result for bfs_pcie:                         7.5798 GB/s
    result for bfs_teps:                      376338000.0000 Edges/s
    result for fft_sp:                        2293.4700 GFLOPS
    result for fft_sp_pcie:                     35.2146 GFLOPS
    result for ifft_sp:                       2279.9500 GFLOPS
    result for ifft_sp_pcie:                    35.2113 GFLOPS
    result for sgemm_n:                       9353.9900 GFLOPS
    result for sgemm_t:                       9233.5700 GFLOPS
    result for sgemm_n_pcie:                  3795.6600 GFLOPS
    result for sgemm_t_pcie:                  3776.0300 GFLOPS
    result for md_sp_flops:                    998.9960 GFLOPS
    result for md_sp_bw:                       765.6000 GB/s
    result for md_sp_flops_pcie:                57.1860 GFLOPS
    result for md_sp_bw_pcie:                   43.8256 GB/s
    result for md5hash:                         31.0101 GHash/s
    result for reduction:                      302.1470 GB/s
    result for reduction_pcie:                  11.8165 GB/s
    result for scan:                            85.5936 GB/s
    result for scan_pcie:                        5.9156 GB/s
    result for sort:                             1.8335 GB/s
    result for sort_pcie:                        1.4230 GB/s
    result for spmv_csr_scalar_sp:              56.2969 Gflop/s
    result for spmv_csr_scalar_sp_pcie:          1.1735 Gflop/s
    result for spmv_csr_scalar_pad_sp:          57.7069 Gflop/s
    result for spmv_csr_scalar_pad_sp_pcie:      1.0237 Gflop/s
    result for spmv_csr_vector_sp:              88.8624 Gflop/s
    result for spmv_csr_vector_sp_pcie:          1.1824 Gflop/s
    result for spmv_csr_vector_pad_sp:          93.5053 Gflop/s
    result for spmv_csr_vector_pad_sp_pcie:      1.0307 Gflop/s
    result for spmv_ellpackr_sp:                55.6213 Gflop/s
    result for stencil:                        623.3620 GFLOPS
    result for triad_bw:                        12.2925 GB/s
    result for s3d:                            447.3900 GFLOPS
    result for s3d_pcie:                       144.7300 GFLOPS
\end{verbatim}

\break

\section{NVIDIA V100 SXM2 SHOC Single Precision Output}
\begin{verbatim}
    result for bspeed_download:                 37.5824 GB/sec
    result for bspeed_readback:                 38.8395 GB/sec
    result for maxspflops:                    15516.8000 GFLOPS
    result for gmem_readbw:                    888.3320 GB/s
    result for gmem_readbw_strided:            479.0020 GB/s
    result for gmem_writebw:                   742.7190 GB/s
    result for gmem_writebw_strided:            59.8676 GB/s
    result for lmem_readbw:                   9453.9000 GB/s
    result for lmem_writebw:                  10179.5000 GB/s
    result for tex_readbw:                    1512.2300 GB/sec
    result for bfs:                             10.5773 GB/s
    result for bfs_pcie:                         7.3547 GB/s
    result for bfs_teps:                      378866000.0000 Edges/s
    result for fft_sp:                        2278.6600 GFLOPS
    result for fft_sp_pcie:                    175.6960 GFLOPS
    result for ifft_sp:                       2260.2600 GFLOPS
    result for ifft_sp_pcie:                   176.1480 GFLOPS
    result for sgemm_n:                       14643.4000 GFlops
    result for sgemm_t:                       14347.2000 GFlops
    result for sgemm_n_pcie:                  8729.9400 GFlops
    result for sgemm_t_pcie:                  8623.8000 GFlops
    result for md_sp_flops:                    912.5020 GFLOPS
    result for md_sp_bw:                       699.3130 GB/s
    result for md_sp_flops_pcie:               132.5160 GFLOPS
    result for md_sp_bw_pcie:                  101.5560 GB/s
    result for md5hash:                         34.7245 GHash/s
    result for nn_learning:                     BenchmarkError
    result for nn_learning_pcie:                BenchmarkError
    result for reduction:                      325.9520 GB/s
    result for reduction_pcie:                  34.3715 GB/s
    result for scan:                           198.6420 GB/s
    result for scan_pcie:                       17.0488 GB/s
    result for sort:                            21.3906 GB/s
    result for sort_pcie:                        9.9923 GB/s
    result for spmv_csr_scalar_sp:              68.7036 Gflop/s
    result for spmv_csr_scalar_sp_pcie:          6.3301 Gflop/s
    result for spmv_csr_scalar_pad_sp:          77.8411 Gflop/s
    result for spmv_csr_scalar_pad_sp_pcie:      6.6088 Gflop/s
    result for spmv_csr_vector_sp:             164.1990 Gflop/s
    result for spmv_csr_vector_sp_pcie:          6.6855 Gflop/s
    result for spmv_csr_vector_pad_sp:         172.6370 Gflop/s
    result for spmv_csr_vector_pad_sp_pcie:      6.9219 Gflop/s
    result for spmv_ellpackr_sp:                90.2127 Gflop/s
    result for stencil:                        690.8430 GFLOPS
    result for triad_bw:                        36.0896 GB/s
    result for s3d:                            462.9920 GFLOPS
    result for s3d_pcie:                       363.5560 GFLOPS
\end{verbatim}

\break

\section{NVIDIA A100 PCIe SHOC Single Precision Output}
\begin{verbatim}
    result for bspeed_download:                 26.8296 GB/sec
    result for bspeed_readback:                 27.1204 GB/sec
    result for maxspflops:                    19309.1000 GFLOPS
    result for gmem_readbw:                   2865.4300 GB/s
    result for gmem_readbw_strided:            601.8820 GB/s
    result for gmem_writebw:                  2214.4100 GB/s
    result for gmem_writebw_strided:           160.3150 GB/s
    result for lmem_readbw:                   14217.0000 GB/s
    result for lmem_writebw:                  15983.8000 GB/s
    result for tex_readbw:                    1560.7800 GB/sec
    result for ocl_kernel:                       0.0004 sec
    result for ocl_queue:                        0.0050 ms
    result for bfs:                             23.4141 GB/s
    result for bfs_pcie:                        13.4664 GB/s
    result for bfs_teps:                      556110000.0000 Edges/s
    result for fft_sp:                        4006.2800 GFLOPS
    result for fft_sp_pcie:                     74.3269 GFLOPS
    result for ifft_sp:                       3941.3600 GFLOPS
    result for ifft_sp_pcie:                    74.3042 GFLOPS
    result for sgemm_n:                       13458.9000 GFLOPS
    result for sgemm_t:                       12928.0000 GFLOPS
    result for sgemm_n_pcie:                  5807.9600 GFLOPS
    result for sgemm_t_pcie:                  5717.3800 GFLOPS
    result for md_sp_flops:                   1558.6600 GFLOPS
    result for md_sp_bw:                      1194.5100 GB/s
    result for md_sp_flops_pcie:                91.7757 GFLOPS
    result for md_sp_bw_pcie:                   70.3341 GB/s
    result for md5hash:                         42.8308 GHash/s
    result for reduction:                      239.9240 GB/s
    result for reduction_pcie:                  23.7619 GB/s
    result for scan:                            74.4782 GB/s
    result for scan_pcie:                       11.3414 GB/s
    result for sort:                             1.8775 GB/s
    result for sort_pcie:                        1.6457 GB/s
    result for spmv_csr_scalar_sp:              82.0996 Gflop/s
    result for spmv_csr_scalar_sp_pcie:          2.3338 Gflop/s
    result for spmv_csr_scalar_pad_sp:          69.8097 Gflop/s
    result for spmv_csr_scalar_pad_sp_pcie:      2.4474 Gflop/s
    result for spmv_csr_vector_sp:             137.9340 Gflop/s
    result for spmv_csr_vector_sp_pcie:          2.3610 Gflop/s
    result for spmv_csr_vector_pad_sp:         145.6470 Gflop/s
    result for spmv_csr_vector_pad_sp_pcie:      2.4929 Gflop/s
    result for spmv_ellpackr_sp:                83.6452 Gflop/s
    result for stencil:                       1303.0500 GFLOPS
    result for triad_bw:                        25.2434 GB/s
    result for s3d:                            827.5860 GFLOPS
    result for s3d_pcie:                       407.1100 GFLOPS
\end{verbatim}

\break

\section{Apple M1 SHOC Single Precision Output}

\begin{verbatim}
    result for bspeed_download:                846.8210 GB/sec
    result for bspeed_readback:                848.3620 GB/sec
    result for maxspflops:                    297487000.0000 GFLOPS
    result for gmem_readbw:                   2773.7900 GB/s
    result for gmem_readbw_strided:           2172.6800 GB/s
    result for gmem_writebw:                  2931.5800 GB/s
    result for gmem_writebw_strided:          1661.5100 GB/s
    result for lmem_readbw:                   25617.3000 GB/s
    result for lmem_writebw:                  35432.5000 GB/s
    result for tex_readbw:                    3091.7900 GB/sec
    result for ocl_kernel:                       0.0000 sec
    result for ocl_queue:                        0.0018 ms
    result for bfs:                             47.7882 GB/s
    result for bfs_pcie:                        46.1705 GB/s
    result for bfs_teps:                      33420200.0000 Edges/s
    result for fft_sp:                        4021.3500 GFLOPS
    result for fft_sp_pcie:                     52.5703 GFLOPS
    result for ifft_sp:                       3983.4400 GFLOPS
    result for ifft_sp_pcie:                    52.5638 GFLOPS
    result for sgemm_n:                       20437.2000 GFLOPS
    result for sgemm_t:                       21263.8000 GFLOPS
    result for sgemm_n_pcie:                  18435.0000 GFLOPS
    result for sgemm_t_pcie:                  19708.6000 GFLOPS
    result for md_sp_flops:                   3706.0700 GFLOPS
    result for md_sp_bw:                      2840.2200 GB/s
    result for md_sp_flops_pcie:              2279.1000 GFLOPS
    result for md_sp_bw_pcie:                 1746.6300 GB/s
    result for md5hash:                         76.5948 GHash/s
    result for reduction:                     2347.8900 GB/s
    result for reduction_pcie:                 807.0470 GB/s
    result for scan:                            19.6674 GB/s
    result for scan_pcie:                       19.0673 GB/s
    result for sort:                             0.3223 GB/s
    result for sort_pcie:                        0.3221 GB/s
    result for spmv_csr_scalar_sp:             103.2920 Gflop/s
    result for spmv_csr_scalar_sp_pcie:         73.4816 Gflop/s
    result for spmv_csr_scalar_pad_sp:         103.0530 Gflop/s
    result for spmv_csr_scalar_pad_sp_pcie:     74.4468 Gflop/s
    result for spmv_csr_vector_sp:             574.9480 Gflop/s
    result for spmv_csr_vector_sp_pcie:        176.4760 Gflop/s
    result for spmv_csr_vector_pad_sp:         605.5290 Gflop/s
    result for spmv_csr_vector_pad_sp_pcie:    185.8990 Gflop/s
    result for spmv_ellpackr_sp:               388.9170 Gflop/s
    result for stencil:                         64.3946 GFLOPS
    result for triad_bw:                       713.6200 GB/s
    result for s3d:                           1590.5500 GFLOPS
    result for s3d_pcie:                      1554.4700 GFLOPS
\end{verbatim}

\break

\section{Apple M1 Ultra SHOC Single Precision Output}
\begin{verbatim}
    result for bspeed_download:               2237.6300 GB/sec
    result for bspeed_readback:               2024.0300 GB/sec
    result for maxspflops:                    655714000.0000 GFLOPS
    result for gmem_readbw:                   10938.8000 GB/s
    result for gmem_readbw_strided:           37415.7000 GB/s
    result for gmem_writebw:                  10237.2000 GB/s
    result for gmem_writebw_strided:          13399.4000 GB/s
    result for lmem_readbw:                   301038.0000 GB/s
    result for lmem_writebw:                  358733.0000 GB/s
    result for tex_readbw:                    28706.6000 GB/sec
    result for ocl_kernel:                       0.0000 sec
    result for ocl_queue:                        0.0027 ms
    result for bfs:                           1009.6800 GB/s
    result for bfs_pcie:                       919.6110 GB/s
    result for bfs_teps:                      73887900.0000 Edges/s
    result for fft_sp:                        37314.0000 GFLOPS
    result for fft_sp_pcie:                    123.9950 GFLOPS
    result for ifft_sp:                       36082.6000 GFLOPS
    result for ifft_sp_pcie:                   123.9810 GFLOPS
    result for sgemm_n:                       248243.0000 GFLOPS
    result for sgemm_t:                       258108.0000 GFLOPS
    result for sgemm_n_pcie:                  188918.0000 GFLOPS
    result for sgemm_t_pcie:                  195467.0000 GFLOPS
    result for md_sp_flops:                   18996.3000 GFLOPS
    result for md_sp_bw:                      14558.1000 GB/s
    result for md_sp_flops_pcie:              12922.9000 GFLOPS
    result for md_sp_bw_pcie:                 9903.7200 GB/s
    result for md5hash:                        620.4690 GHash/s
    result for reduction:                     7491.7000 GB/s
    result for reduction_pcie:                4660.4200 GB/s
    result for scan:                            57.8603 GB/s
    result for scan_pcie:                       57.3959 GB/s
    result for sort:                             0.6722 GB/s
    result for sort_pcie:                        0.6721 GB/s
    result for spmv_csr_scalar_sp:            1054.3300 Gflop/s
    result for spmv_csr_scalar_sp_pcie:        620.2710 Gflop/s
    result for spmv_csr_scalar_pad_sp:        1074.9700 Gflop/s
    result for spmv_csr_scalar_pad_sp_pcie:    747.3490 Gflop/s
    result for spmv_csr_vector_sp:            2657.9100 Gflop/s
    result for spmv_csr_vector_sp_pcie:        961.8070 Gflop/s
    result for spmv_csr_vector_pad_sp:        2916.4600 Gflop/s
    result for spmv_csr_vector_pad_sp_pcie:   1332.3300 Gflop/s
    result for spmv_ellpackr_sp:               694.4130 Gflop/s
    result for stencil:                        432.3320 GFLOPS
    result for triad_bw:                      1614.1800 GB/s
    result for s3d:                           11276.0000 GFLOPS
    result for s3d_pcie:                      10913.0000 GFLOPS
\end{verbatim}

\end{document}